\documentclass[conference]{IEEEtran}

\IEEEoverridecommandlockouts 

\usepackage{cite}
\usepackage{amsmath,amssymb,amsfonts}
\usepackage{algorithmic}
\usepackage{graphicx}
\usepackage{booktabs}
\usepackage{multicol}
\usepackage{makecell}
\usepackage{xcolor}
\usepackage{flushend}
\usepackage[super]{nth}
\usepackage{enumitem}
\usepackage{epstopdf}
\usepackage{enumitem}
\usepackage[warn]{textcomp}
\usepackage{orcidlink}

\def\conv8k{{8\textrightarrow48\,kHz}}
\def\48k{{48\,kHz}}
\def\8k{{8\,kHz}}
\def\tablefont{\small}

\makeatletter
\def\bstctlcite{\@ifnextchar[{\@bstctlcite}{\@bstctlcite[@auxout]}}
\def\@bstctlcite[#1]#2{\@bsphack
  \@for\@citeb:=#2\do{%
    \edef\@citeb{\expandafter\@firstofone\@citeb}%
    \if@filesw\immediate\write\csname #1\endcsname{\string\citation{\@citeb}}\fi}%
  \@esphack}
\makeatother
\bstctlcite{IEEEexample:BSTcontrol}  

\begin{document}
\title{Sampling Frequency Independent Dialogue Separation}
\author{\IEEEauthorblockN{Jouni Paulus\IEEEauthorrefmark{1}~\orcidlink{0000-0003-2283-2062}\thanks{\IEEEauthorrefmark{1}Now with WS Audiology, Erlangen, Germany.}}
\IEEEauthorblockA{\textit{Fraunhofer Institute for Integrated Circuits IIS}, and \\
\textit{International Audio Laboratories Erlangen\IEEEauthorrefmark{2}}\thanks{\IEEEauthorrefmark{2}International Audio Laboratories Erlangen is a joint institution of Universit{\"a}t Erlangen-N{\"u}rnberg and Fraunhofer IIS.}\\
Erlangen, Germany}%
\and
\IEEEauthorblockN{Matteo Torcoli~\orcidlink{0000-0003-2834-9194}}
\IEEEauthorblockA{\textit{Fraunhofer Institute for Integrated Circuits IIS}, and \\
\textit{International Audio Laboratories Erlangen\IEEEauthorrefmark{2}} \\
Erlangen, Germany}}%

\maketitle
\begin{abstract}  
In some DNNs for audio source separation, the relevant model parameters are independent of the sampling frequency of the audio used for training.
Considering the application of dialogue separation, this is shown for two DNN architectures: a U-Net and a fully-convolutional model.
The models are trained with audio sampled at 8\,kHz.
The learned parameters are transferred to models for processing audio at 48\,kHz.
The separated audio sources are compared with the ones produced by the same model architectures trained with 48\,kHz versions of the same training data.
A listening test and computational measures show that there is no significant perceptual difference between the models trained with 8\,kHz or with 48\,kHz.
This transferability of the learned parameters allows for a faster and computationally less costly training.
It also enables using training datasets available at a lower sampling frequency than the one needed by the application at hand, or using data collections with multiple sampling frequencies.
\end{abstract}
%

\section{Introduction}
\label{sec:intro}
When developing and training a deep neural network (DNN) for audio signal processing, the input signal sampling frequency is normally fixed and defined by the target application, e.g., 8 or 16\,kHz for speech, 44.1\,kHz for music, and 48\,kHz for broadcast applications.
Alternatively, the sampling frequency of the model is dictated by the available training data.
If we want to process signals at a sampling frequency different from the one in training, signal re-sampling may be needed, a new model for the new sampling frequency needs to be trained, or a different paradigm needs to be applied, e.g.,~\cite{Subramani21-WASPAA}.

Focusing on the dialogue separation application~\cite{Paulus19-JAES, Torcoli21-IBC}, this paper presents the observation of sampling frequency independence of relevant model parameters in certain audio source separation DNN architectures.
In these architectures the trainable parameters are effectively independent of the actual signal sampling frequency.
This opens the possibilities for training a single model with  multiple datasets of different sampling frequencies, or training and deploying a model in different sampling frequencies.
Additionally, training at a lower sampling frequency is computationally cheaper and faster, possibly reducing the carbon footprint.

To the best of our knowledge, the only directly related work is by Saito et al.~\cite{Saito21-EUSIPCO}.
They construct a sampling-frequency-independent (SFI) convolutional analysis and synthesis filter bank (i.e., encoder and decoder) for ConvTasNet~\cite{Luo19-TASLPACM} by sampling analog gammatone filter impulse responses at the provided sampling frequency.
The filter center frequencies are independent of the sampling frequency, and their number is fixed, meaning that the encoded representation covers always the same frequency range.
In this contribution, we show that it is possible to achieve sampling frequency independence also with the more commonly-used short-time Fourier transform (STFT), with the advantage that the analytic definition of the transform allows covering the full frequency range.
We propose that the key is keeping the spectral and temporal granularity constant and independent of the sampling frequency.

In the remainder of this paper, we look into two DNN architectures built into a common framework (Sec.~\ref{sec:method}), we describe an experimental setup for the sampling frequency change (Sec.~\ref{sec:variants}), and show that the parameters of these models can be transferred from a model trained with 8\,kHz data to a model for processing 48\,kHz data with no significant quality degradation compared to a native 48\,kHz model (Sec.~\ref{sec:eval}).

\section{Method}
\label{sec:method}
\subsection{Sampling frequency independence}
\label{sec:sfi}
\begin{figure}[tb]
\centering
\centerline{\includegraphics[width=0.75\columnwidth]{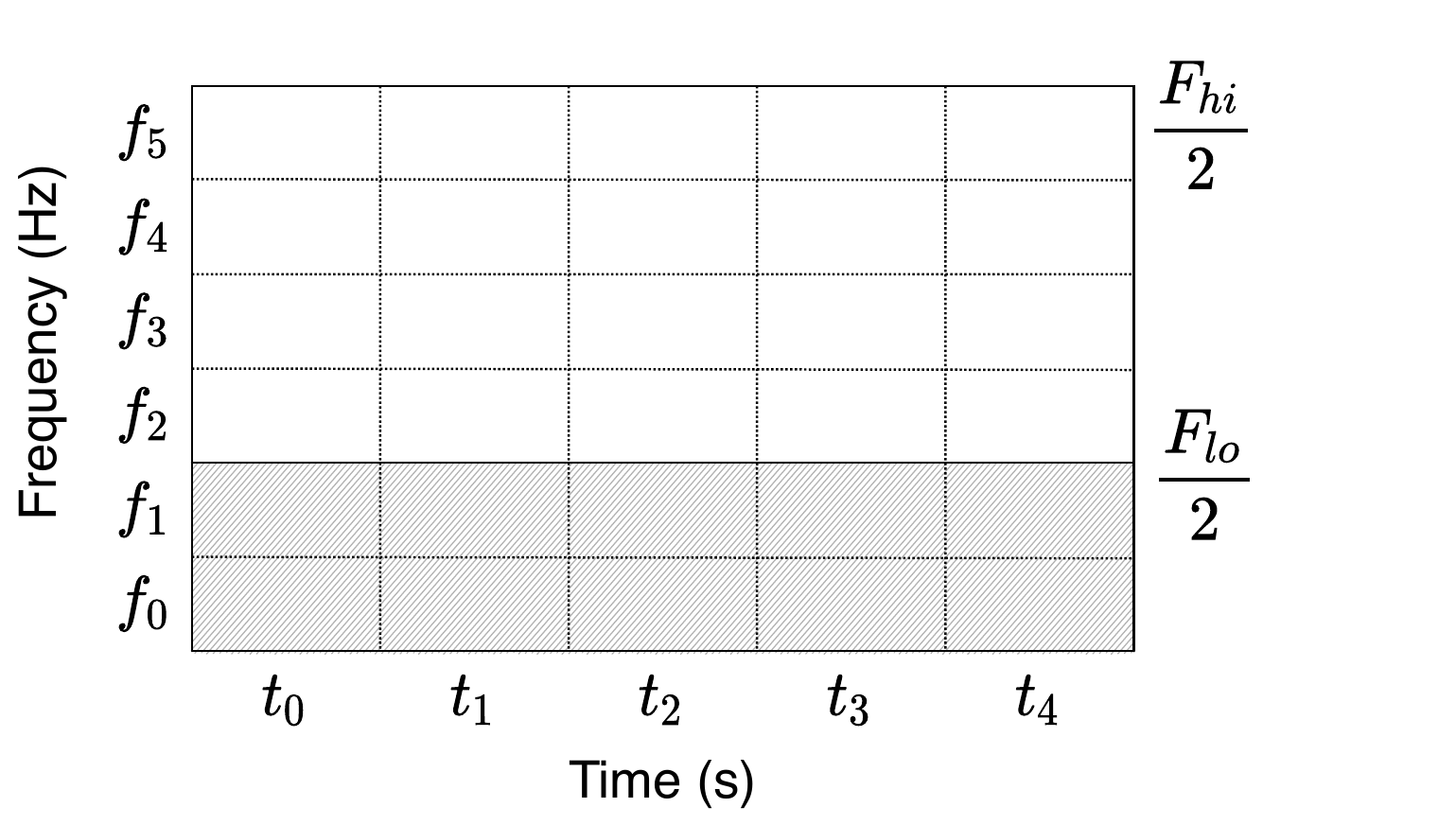}}
\caption{Time/frequency tiles after the \emph{encoding} layer with an STFT adapted to two different sampling frequencies $F_{lo}$ and $F_{hi} = 3F_{lo}$. The shaded tiles represent the resolution obtained for the lower sampling frequency.
}
\label{fig:tfgrid}
\end{figure}

The main assumption in this paper is that the time-frequency resolution of the representation in the DNN is constant even when the signal sampling frequency is different.
Let us consider the STFT representation as an example.
A constant time-frequency resolution across sampling frequencies can be fulfilled by setting a constant frame length and spacing in units of seconds.
From this and the sampling frequency, the length of the transform can be determined.
As a result, the STFT representations obtained from two signals with different sampling frequencies (but same duration in seconds) will have the same number of frames, e.g., $t_0, \ldots, t_4$, as illustrated in Fig.~\ref{fig:tfgrid}.
The number of frequency axis elements (or \emph{bins}) will differ, e.g., $f_0, f_1$ for the lower sampling frequency, and $f_0, \ldots, f_5$ for a sampling frequency 3 times higher.
Still, the spacing of the bins is the same, i.e., $f_0$ and $f_1$ will refer to the same frequency sub-bands at both sampling frequencies.

Operations acting only on a local subset of the frequency axis elements (e.g., $f_0$ and $f_1$) will now have similar information available on the overlapping (low) frequency region of all sampling frequencies.
Examples of such operations are 1D and 2D convolutions, or 1D recurrent layers operating along the frequency axis or per sub-band, e.g.,~\cite{Li19-WASPAA}.
Since operations of these layers are independent of the absolute bin frequency, it appears that training with band-limited data will result into parameters that will work also for the frequency region not present in the training.

This is not a generic solution, but dependent on the exact network architecture and layers used, e.g., a model containing a fully-connected layer cannot be trivially converted.

One should note that even though the sampling frequencies used for the main demonstration in this paper have an integer multiple relationship, this is not a strict requirement in the real world.
In fact, the resulting frame lengths and transform lengths used in the experiments are not exact integer multiples (see Sec.~\ref{sec:dnn:common}).
Even though the time/frequency-resolutions are not identical and the assumption outlined above is relaxed, no significant perceptual quality degradation is observed (Sec.~\ref{sec:eval}).
As a related effect, the resulting time-to-frequency transform lengths may be non-power-of-2. 
\subsection{Common DNN framework}
\label{sec:dnn:common}
\begin{figure}[tb]
\centering
\centerline{\includegraphics[width=\columnwidth]{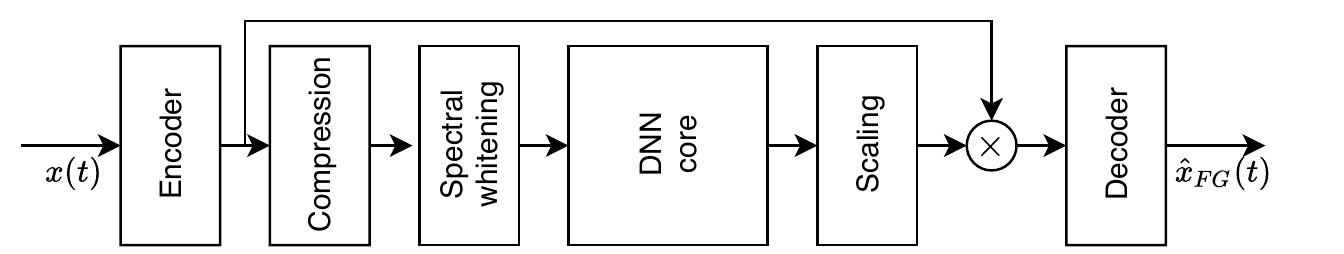}}
\caption{The common DNN framework used in the experiments.
The block \emph{DNN core} is either a U-Net or a convolutional block, depending on the tested system.}
\label{fig:framework}
\end{figure}

The experiments make use of a U-Net core~\cite{Ronneberger15-MICCAI, Jansson17-ISMIR, Pretet19-ICASSP, Hennequin20-JOSS} and a fully-convolutional core, both inside the encoder-decoder architecture shown in Fig.~\ref{fig:framework} with the common processing steps consisting of:
\begin{enumerate}[leftmargin=0pt, labelwidth=-2em,font=\itshape]
\item \emph{Encoder:} The time-domain input signal $x(t)$ is transformed into an encoded representation. The input is assumed to be a mixtures of a foreground target and background non-target components $x(t) =\nolinebreak x_{FG}(t)+x_{BG}(t)$.
We use STFT implemented as a strided 1D convolution.
The number of convolutional filters depends on the sampling frequency for which the model is instantiated.
Each filter corresponds to the combination of the windowing function (here, sine window) and the time-reversed impulse response of the STFT kernel.
The frame length is ca. 42.7\,ms, corresponding to 2048 samples at 48\,kHz sampling frequency (342 samples at 8\,kHz and 1882 samples at 44.1\,kHz), and the stride is 50\%.
\item \emph{Compression:} The magnitudes of the spectral elements $c =\nolinebreak \Re(c) + i\Im(c)$ are compressed with $c_{z} =\nolinebreak q\Re(c) + qi\Im(c)$, where $q$ is the effective magnitude scaling computed as $q =\nolinebreak \log{(\alpha+|c|)} / |c|$, with $\alpha=1$.
\item The compressed values are split into the real and imaginary parts and stacked from all input audio channels into the \emph{channels} dimension of the network tensors.
\item \emph{Spectral whitening:} For each frequency bin, subtract the mean and divide by standard deviation over the training data.
\item \emph{DNN core:} Separation filters (masks) are produced for cross-channel filtering.
\item \emph{Scaling:} Global scalar offset and scaling are applied to the filters to allow exceeding the range of the last activation.
\item The separation filters are applied on the original encoded representation.
\item \emph{Decoder:} The encoded result is transformed back to time domain producing an estimate of the target signal $\hat{x}_{FG}(t)$.
We use a convolutional STFT synthesis filter bank.
\item An estimate of the non-target component is obtained by a subtraction from the input mixture: $\hat{x}_{BG}(t) = x(t) - \hat{x}_{FG}(t)$.
\end{enumerate}

Applying the proposed sampling frequency conversion for the models implemented in this framework is straightforward.
First, the frame length is adjusted according to sampling frequency and the filters of the STFT-based encoder and decoder are designed for the new length.
After the encoder, the data is of the same form with only a different number of elements in the dimension corresponding to frequency.
The spectral whitening layer needs to be assigned with new per-frequency parameters, which can be computed, e.g., with one pass over target mixtures.
As the remaining operations fulfil the assumption of being agnostic to the absolute frequency and apply their function to the whole input regardless of its size, the conversion is now ready.

\subsection{U-Net}
\label{sec:dnn:unet}
The U-Net core follows closely the structure of Spleeter~\cite{Hennequin20-JOSS} with the following changes.
The up-sampling path uses factor 2 nearest-neighbor up-sampling, followed by a $5 \times 5$ convolution, instead of the transposed strided convolutions used in the original model.
The number of filters in the last up-sampling convolution and the output $5 \times 5$ convolution are equal to the number of separation filters.
The output activation is tanh.

\subsection{Convolutional network (CNN)}
\label{sec:dnn:cnn}
The second DNN core used in the experiments is a fully-convolutional network developed for dialogue separation~\cite{Paulus19-JAES}.
Variants of this model have been tested in earlier investigations~\cite{Torcoli21-IBC, Strauss21-INTERSPEECH, Torcoli21-WASPAA}.
The model consists of a stack of 24 convolutional blocks, each having a frequency-domain padding in reflection mode, a $3 \times 5$ (time $\times$ frequency) 2D convolution with 32 filters, a ReLU activation, and a layer normalization for the \emph{channels}.
The last block uses tanh activation and the number of filters is equal to the number of separation filters.

\section{Model Variants}
\label{sec:variants}

Our experiments make use of the two core architectures in these variants:
\begin{itemize}[leftmargin=0pt, labelwidth=-2em,font=\itshape]
\item \emph{\48k}: This is the reference condition, in which the model is both trained with and applied on 48\,kHz data.
\item \emph{\8k}: The model is both trained with and applied on 8\,kHz data.
The test items are down-sampled for processing, and the model outputs are up-sampled back to 48\,kHz.
\item \emph{$F_{train}$\textrightarrow$F_{test}$\,kHz}: The model is trained with data having sampling frequency $F_{train} =$ 8\,kHz and the parameters are copied to a model for data with sampling frequency of \mbox{$F_{test} \in$ \{44.1, 48\}\,kHz}.
The spectral whitening layer parameters are obtained from the training mixtures at $F_{test}$\,kHz. Note that $F_{train}$ and $F_{test}$ do not necessarily have an integer ratio.
\end{itemize}

\subsection{Data}
The data used in our experiments consists of almost 21 hours of stereo audio from real broadcast content.
All items have foreground signals (speech or dialogue) and matching background signals (music, effects, non-speech).
The native sampling frequency of the data is 48\,kHz, a full copy of the data is down-sampled to 8\,kHz.
The training data (14\,h\,18\,min) and the test set (1\,h) are independent and the same as in~\cite{Strauss21-INTERSPEECH}.
The validation data consists of 5\,h\,36\,min of audio from the same item pool as the training data.
The training examples are used in full length.
Online data augmentation is used for each example with a random offset in the beginning (max.\,10\,ms) to avoid identical time ranges, 33\% chance of being down-mixed into mono (in stereo experiments), a random gain in the range $[-6,+6]$\,dB, and a signal-to-background mixing ratios modification by $[-6,+6]$\,dB.
The order of the training examples and the augmentation are randomized in each epoch.

The main experiments use models for separating stereo components from stereo input signals.
For verifying that the models do not rely only on spatial information, the computational evaluation is performed additionally for models separating mono components from mono inputs.
For the mono tests, the entire data set is converted to mono and the down-mixing augmentation is disabled.

\subsection{Training}
We use time-domain mean absolute error (MAE) as loss and ADADELTA~\cite{Zeiler12-arxiv} as the optimizer.
The training is run until the validation loss does not improve in 10 epochs, and the model with the lowest validation loss is used in the evaluation.

The stereo training of the U-Net converged on epochs 96 and 85 for the 8\,kHz and 48\,kHz models, and of the CNN on epochs 28 and 30 for the 8\,kHz and 48\,kHz models (for mono, epochs 61 and 136 for the U-Net and 41 and 43 for the CNN).
The number of trainable parameters in the stereo models with the U-Net core is 9,827,330, and with the CNN core 359,438 and does not depend on the sampling frequency.
These numbers exclude the parameters of the spectral whitening operation.
On our system, the average per-epoch training duration for the U-Net core is 42\,s for the 8\,kHz version and 6\,min\,45\,s for the 48\,kHz version, meaning 9.6 times faster training on the lower sampling frequency data.
For the CNN, the average per-epoch training durations are 6\,min and 42\,min, meaning 7 times faster training with the lower sampling frequency.

\begin{figure}[tb]
\centering
\includegraphics[width=0.9\columnwidth]{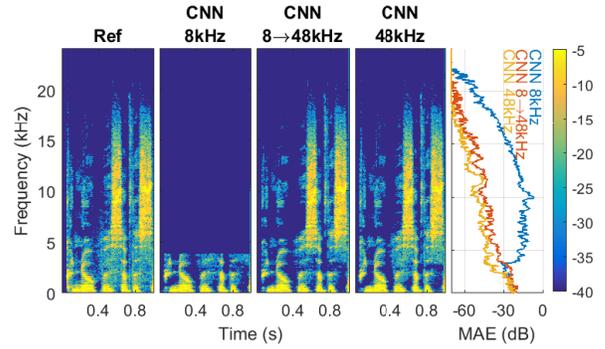}
\caption{Examples magnitude spectrograms of a 1\,s excerpt.
The frequency-dependent MAE is between the reference magnitude spectrum \emph{Ref} and the speech estimates that the CNN variants separated from the input mixture (input SI-SIR = 8.5\,dB).}
\label{fig:spectra}
\end{figure}

\section{Evaluation}
\label{sec:eval}
Fig.~\ref{fig:spectra} visualizes example magnitude spectra of the outputs of the three CNN model variants.
While obviously \emph{CNN\,\8k} does not have any content above 4\,kHz, both \emph{CNN\,\conv8k} and \emph{CNN\,\48k} are able to separate the target with little errors in high band.
In particular, \emph{CNN\,\conv8k} exhibits no major difference with  \emph{CNN\,\48k} in high band separation performance, even if that frequency range was not seen during training.
A more formal evaluation follows.

\begin{table*}[t!]
\centering
\caption{Computational evaluation of stereo models.
The rows $\Delta$SI-SDR, $\Delta$SI-SIR, and $\Delta$2f show the mean per-item improvement and the standard deviation for the given measure with respect to the input mixture.
The row SI-SAR shows the mean per-item absolute value and the standard deviation of the processed conditions.
The input mixtures were artifact-free.
The values for the input mixture were SI-SDR~6.5$\pm$7.0\,dB, SI-SIR~6.5$\pm$7.0\,dB, and 2f-model~20.6$\pm$11.9.
}
\tablefont
\begin{tabular}{l|rrrrrrrr}
                    & \makecell{CNN\\\8k} & \makecell{CNN\\\conv8k} & \makecell{CNN\\8\textrightarrow44.1\,kHz}& \makecell{CNN\\\48k} & \makecell{U-Net\\\8k} & \makecell{U-Net\\\conv8k} & \makecell{U-Net\\8\textrightarrow44.1\,kHz} & \makecell{U-Net\\\48k} \\
\midrule
$\Delta$SI-SDR (dB) &   3.0$\pm$5.3       & \bf{8.3$\pm$4.1}        &  \bf{8.3$\pm$4.0}                        & \bf{8.2$\pm$4.3}     &  2.4$\pm$5.0          &  6.9$\pm$4.1              &     5.9$\pm$4.3                             & 7.2$\pm$4.5            \\
$\Delta$SI-SIR (dB) &  \bf{18.5$\pm$8.4}  &  \bf{18.9$\pm$8.8}      &  \bf{18.7$\pm$8.8}                       & 17.7$\pm$8.4         &  15.8$\pm$8.0         &  16.5$\pm$8.4             &    17.2$\pm$9.5                             & 16.7$\pm$7.4           \\
SI-SAR (dB)         &  9.9$\pm$4.2        &  \bf{15.6$\pm$5.9}      &  \bf{15.6$\pm$5.9}                       & \bf{15.8$\pm$6.1}    &  9.4$\pm$4.5          &  14.5$\pm$6.5             &    13.2$\pm$6.0                             & 14.6$\pm$6.3           \\
$\Delta$2f          &  3.8$\pm$8.3        &  15.7$\pm$6.7           &  15.8$\pm$6.7                            & \bf{16.6$\pm$7.0}    &  3.8$\pm$8.2          &  14.1$\pm$7.3             &    11.2$\pm$7.5                             & 13.1$\pm$7.6           \\
\end{tabular}
\label{table:results:stereo}
\end{table*}

\begin{table*}[t!]
\centering
\caption{Computational evaluation of mono models.
See the caption of Table~\ref{table:results:stereo} for the description of the rows.
The values for the input mixture were SI-SDR~7.7$\pm$7.0\,dB, SI-SIR~7.7$\pm$7.0\,dB, and 2f-model~20.8$\pm$12.0.
}
\tablefont
\begin{tabular}{l|rrrrrr}
                    & \makecell{CNN\\\8k} & \makecell{CNN\\\conv8k} & \makecell{CNN\\\48k} & \makecell{U-Net\\\8k} & \makecell{U-Net\\\conv8k} & \makecell{U-Net\\\48k} \\
\midrule
$\Delta$SI-SDR (dB) &   1.6$\pm$5.1       &  \bf{6.6$\pm$3.7}       &  \bf{6.8$\pm$3.8}    &  1.0$\pm$4.8          &  5.3$\pm$3.8              &  5.7$\pm$4.2\\
$\Delta$SI-SIR (dB) &  14.1$\pm$6.8       &  14.7$\pm$7.2           &  \bf{15.4$\pm$7.2}   &  12.7$\pm$7.5         &  13.6$\pm$7.8             &  12.7$\pm$7.4\\
SI-SAR (dB)         &   9.8$\pm$4.2       &  \bf{15.5$\pm$6.1}      &  \bf{15.5$\pm$6.0}   &  9.4$\pm$4.5          &  14.5$\pm$6.5             &  \bf{15.2$\pm$6.6}\\
$\Delta$2f          &  -1.5$\pm$8.2       &  9.0$\pm$5.8            &  \bf{10.4$\pm$5.7}   &  -1.4$\pm$8.6         &  8.4$\pm$6.7              &  \bf{10.1$\pm$6.0}\\
\end{tabular}
\label{table:results:mono}
\end{table*}

\subsection{Computational evaluation}
\label{sec:computational}
The computational evaluation uses the scale-invariant signal-to-distortion (SI-SDR), signal-to-artifacts (SI-SAR), and signal-to-interference (SI-SIR)~\cite{Vincent06-TASLP, LeRoux19-ICASSP} evaluated on the separated foreground signal with the 44.1\,kHz or 48\,kHz reference signal.
Additionally, a measure intended for basic audio quality evaluation by predicting the result of a MUSHRA~\cite{MUSHRA15} listening test, the 2f-model~\cite{Kastner19-WASPAA, Torcoli21-TASLPACM} is computed.
The measures are computed for both the input mixture signal and the model output in order to compute the change in the measure resulting from the models' processing.
The results of the stereo experiments are given in Table~\ref{table:results:stereo} and of the mono experiments in Table~\ref{table:results:mono}.

\subsection{Listening test}
\label{sec:subjective}
A multi-stimulus listening test~\cite{MUSHRA15} is conducted to evaluate the perceptual quality.
The test items are the same as in~\cite{Strauss21-INTERSPEECH}, taken from the stereo test item pool, and they consist of speech on top of various backgrounds.
Each item is 8\,s long.
The separated components are mixed to simulate a dialogue enhancement application by attenuating the background estimate 20\,dB.
The \emph{hidden reference} is obtained from the original component signals used to create the input mixtures, and the \emph{low anchor} is a 4\,kHz low-pass filtered version of this.
Since this reflects the maximum quality a model operating on 8\,kHz signals could reach, the outputs from the \emph{\8k} variants are omitted from the test.
The test conditions are \emph{\conv8k} and \emph{\48k} for both U-Net and CNN cores.

\begin{figure*}[tb]
\centering
\includegraphics[width=1.6\columnwidth]{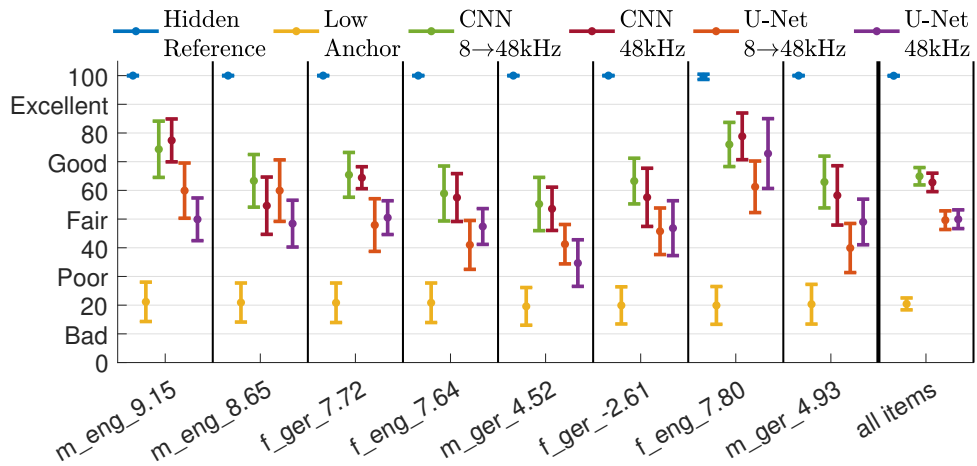}
\caption{Listening test results shown with 95\% confidence intervals (Student's t-distribution, 12 listeners) per item and test condition, and overall on the right.
The items include female (f) and male (m), as well as German (ger) and English (eng) samples, and different input SI-SIR (reported at the end of the each item name).
The constant order of the conditions is shown by the legend. Best viewed in colors.}
\label{fig:mushra}
\end{figure*}

The task given to the listeners was: \emph{"... to rate the overall quality of the conditions in comparison with the given reference."}
The test was taken by 12 expert listeners in their offices using their own high-quality headphones, and no result was post-screened.
Figure~\ref{fig:mushra} shows the test results.

\section{Discussion}
\label{sec:discuss}
The performance differences between the variants \emph{\48k} and \emph{\conv8k} are minimal in all computational measures as well as in the listening test results.
Considering the variance across items and the confidence intervals, no significant difference can be observed between the two variants.
This observation is valid for both U-Net and CNN cores, and for both mono and stereo models.
On the other hand, it is clear especially in the 2f-model output that the \emph{\8k} variants perform worse than the models for 48\,kHz audio.

The results for the 44.1\,kHz sampling frequency target data in Table~\ref{table:results:stereo} show that for the CNN core the conversion \textit{8\textrightarrow44.1\,kHz} works equally well as \emph{\conv8k}, despite the sampling frequencies not having an integer ratio.
The U-Net core architecture converted to 44.1\,kHz shows slight differences in the computational evaluation suggesting a slightly better absolute separation performance, but worse artifact-related performance than for the models processing 48\,kHz data.
A possible reason for this difference may be related to the pooling and un-pooling operations in the U-Net.

Especially from the listening test (Figure~\ref{fig:mushra}), it is clear that the relevant model parameters learned from data with low sampling frequency can be transferred to a model for processing data on considerably higher sampling frequency.
This property is valid for networks using layers that are frequency-axis agnostic, i.e., apply the same operation regardless of the absolute frequency.
Considering ConvTasNet~\cite{Luo19-TASLPACM}, the internal 1D convolutions use the dimension corresponding to frequency as the \emph{channels} dimension, and a different number of frequency bins would result into a different number of filter channels preventing this trivial conversion.
Considering OpenUnmix~\cite{Stoter19-JOSS}, the fully-connected layer across frequencies prevents trivial parameter transfer across sampling frequencies.

In the experiments, the statistics for the spectral whitening layer were learned from one pass through the mixture signals in the training.
In practice, a smaller data set could be used for this, or one could use online estimation. We wanted to eliminate this factor from the evaluation and used the true statistics of the training data.

We used STFT filter bank for demonstrating the observation, since it has an analytic form across sampling frequencies and transform lengths.
Possibly also other filter banks can be similarly used as long as the spacing in the time/frequency grid remains constant, as discussed in Sec.~\ref{sec:sfi}.
Finally, the application of dialogue separation was considered, for which speech is the target signal, which is particularly rich below 4\,kHz.
The transferability of the learned parameters might work to different extents in other source separation tasks with different target types.
This should be investigated in future.

\section{Conclusions}
\label{sec:conclusions}
We have observed that in some DNN architectures for audio source separation, the relevant model parameters are independent of the sampling frequency of the audio.
This allows, e.g., using training data with multiple sampling frequencies for training a single model, or training a model with one sampling frequency and using it on another.
In an experiment, we transferred the parameters from a model trained with 8\,kHz sampling frequency data to a model for processing 48\,kHz sampling frequency data, and no significant perceptual quality difference was observed with respect to a native 48\,kHz model. The training speed was up to 7-10 times faster, depending on the core DNN architecture.

\bibliographystyle{IEEEtran}  
\bibliography{references}

\end{document}